# Influence of phonon and electron excitations on the free energy of defect clusters in solids: A first-principles study

M. Posselt[a)], D. Murali[1], and M. Schiwarth[2]

Helmholtz-Zentrum Dresden - Rossendorf,

Bautzner Landstraße 400, 01328 Dresden, Germany

Although many processes of nanostructure evolution in solids occur at elevated temperatures, basic data obtained from ground state energetics are used in the modeling of these phenomena. In order to illustrate the effect of phonon and electron excitations on the free binding energy of defect clusters, first-principles calculations are performed for vacancy-solute pairs as well as vacancy and Cu dimers, trimers, and quadromers in bcc Fe. Based on the equilibrium atomic positions determined by the relaxation of the supercell with the defect in the ground state under constant volume (CV) as well as zero pressure (ZP) conditions, the contribution of phonon excitations to the free binding energy is calculated within the framework of the harmonic approximation. The contribution of electron excitations is obtained using the corresponding ground state data for the electronic density of states. Quasi-harmonic corrections to the ZP-based results do not yield significant changes in the temperature range relevant for applications. At 1000 K the maximum decrease/increase of the ZP-based data for the absolute value of the free binding energy with respect to the corresponding ground state value is found for the vacancy-W (43%) / vacancy-Mn (35%) pair. These results clearly demonstrate that contributions of phonon and electron excitation to the free binding energy of the defect clusters are generally not negligible. The general behavior of the free binding energy of vacancy and Cu dimers, trimers and quadromers is similar to that of the vacancy-solute pairs. The results obtained in this work are of general importance for studies on the thermodynamics and kinetics of defect clusters in solids.

[a]Author to whom correspondence should be addressed. Electronic mail: m.posselt@hzdr.de

[1]Present address:

Department of Physics, Indian Institute of Technology Madras,

Chennai-600036, India

[2]Present address:

Polymer Competence Center Leoben GmbH, Roseggerstraße 12,

A-8700 Leoben, Austria

Key words: Free energy; Defect cluster, bcc Fe; DFT calculations

# 1. Introduction

The nanostructure of solids has an important influence on their thermal, mechanical, electrical and magnetic properties. Thermodynamics and kinetics of point defects (vacancies, self-interstitials and foreign atoms) and defect clusters determine the evolution of the nanostructure at elevated temperatures. Multiscale modeling can substantially contribute to a better understanding of these processes. For this purpose atomistic rigid-lattice and object kinetic Monte Carlo simulations are employed. The equilibrium state of nanostructures and the phase diagrams of alloys and compounds are often investigated by Metropolis Monte Carlo simulation. The theoretical methods mentioned above require input data which are mainly obtained from first-principle calculations. In most previous studies only ground state properties of point defects and defect clusters determined by Density-Functional-Theory-(DFT)-methods were used to obtain these data. However, at elevated temperatures excitations of phonons, electrons, magnons and other quasi-particles occur so that not the energetics in the ground state but the full thermodynamics, i.e. the free energy of the defects, must be considered. While some authors have pointed out the importance of considering full DFT-based free formation and migration energies of point defects at nonzero temperatures [1-17] only few studies on the free energy of defect clusters have hitherto been published. The work of Yuge *et al.* [18] is probably the only full DFT-based study which is exclusively related to the free energy of larger embedded defect clusters. These authors found a considerable influence of phonon excitations on the nucleation free energy of Cu clusters in bcc Fe. This quantity is closely related to the free binding energy. In investigations on the solubility of foreign atoms the vibrational contribution to the free formation energy of Cu-Cu dimers in bcc Fe and Fe-Fe dimers in fcc Cu was determined by DFT [2]. Such a method was also applied to calculate the free binding energy of vacancy-solute pairs in studies on the vacancy mechanism of solute diffusion in bcc Fe, fcc Al, and fcc Ni (Refs. [3,5,6,19]). In general one can expect that phonon and other excitations may have non-negligible effects on the total free binding energy of embedded defect clusters and on the free binding energy of point defects to these clusters, both of which are crucial input parameters of kinetic and Metropolis Monte Carlo simulations. The replacement of ground state values of the defect binding energy by corresponding free energy data may alter results of the simulations quantitatively and qualitatively. In a fully thermodynamics-based method the migration barriers must be determined by the difference between the free energy at the saddle point and that in the related stable configuration, i.e. the free migration energy must be calculated. However, in state-of-the-art kinetic Monte Carlo simulations migration barriers determined by probing the DFT-energy landscape in the ground state are employed. The corresponding attempt frequencies are mostly set equal to a characteristic vibrational frequency of the solid. In some more detailed DFT-based investigations, in particular on self-diffusion by the vacancy mechanism (cf. Refs. [3,4,15-17,20,21]), contributions of phonon and/or electron excitations to the free migration energy were considered. Also in recent DFT studies on the diffusion of foreign atoms or impurities by the interstitial, the vacancy or the dumbbell mechanism (cf. Refs. [3,5,6,19,22-24]) these contributions were partially taken into account. However, in other recent papers on these diffusion mechanisms the barriers obtained from the DFT-energy landscape in the ground state were used in combination with a constant attempt frequency, i.e. without considering phonon and electron excitations (cf. Refs. [25-29]). It should be also mentioned that several authors clearly demonstrated that the effect of phonon excitations must be considered in first-principles-based calculations of phase diagrams of binary and ternary metallic and non-metallic alloys [30-32]. In these studies DFT methods were applied to supercells containing two or three different atomic species and the free energy of these systems was calculated. From the results temperature-dependent parameters were derived and used in a cluster expansion model for the free energy of larger systems on a rigid lattice.

Finally, for a given composition and temperature the equilibrium state of the compound or alloy was determined using rigid-lattice Metropolis Monte Carlo simulations, which yields the phase diagram.

This paper deals with the influence of phonon and electron excitations on the free energy of defect clusters in bcc Fe. The focus is on defect dimers since from their free binding energy temperature-dependent parameters may be derived, which describe the interaction between pairs of foreign atoms or point defects within the framework of a rigid lattice model. So far, these parameters were obtained from ground-state dimer binding energies. Pair interaction parameters are very important input data of rigid-lattice kinetic and Metropolis Monte Carlo simulations since in most previous studies the total energy of a system was modeled by a sum over first and second nearest neighbor pair interaction terms, cf. Refs. [33-38]. Furthermore, the knowledge of the free binding energy of pairs formed by a vacancy and a foreign atom is required in detailed investigations on the vacancy mechanism of impurity diffusion, cf. Refs. [3,5,6,19,24,39]. As already mentioned above, many previous investigations used the ground-state binding energy instead of the free binding energy. In the present work the ground state energetics as well as the vibrational and electronic contributions to the free binding energy of the divacancy, the Cu-Cu dimer in bcc Fe, as well as dimers formed by the vacancy (v) and the foreign atoms v, C, N, O, Al, Si, Ti, V, Cr, Mn, Co, Ni, Cu, Y, Mo, and W are calculated. Furthermore, the free binding energy of selected trimers ($v_3$, $Cu_3$) and quadromers ($v_4$, $Cu_4$) in bcc Fe is determined. Besides general thermodynamic and kinetic aspects, present investigations shall contribute to a better understanding and an improved modeling of nanostructure evolution in ferritic Fe and Fe-Cr alloys which are important for practical applications, in particular as basic structural materials for present and future nuclear fission and fusion reactors where the relevant temperatures are between about 600 and 1000 K.

## 2. Calculation methodology

### 2.1. Ground state energetics

The DFT calculations were performed with the Vienna ab-initio simulation package VASP [40,41] using plane wave basis sets and pseudopotentials generated within the projector-augmented wave (PAW) approach. The exchange and correlation effects were modeled by the Perdew-Burke-Ernzerhof (PBE) parameterization [42] of the generalized gradient approximation (GGA). In all calculations the spin polarized formalism was applied and a plane wave cutoff of 500 eV was used. The Brillouin zone sampling was performed employing the Monkhorst-Pack scheme [43]. For the integration in the reciprocal space the Methfessel-Paxton smearing method method [44] was applied with a width of 0.2 eV. The calculations were carried out for bcc-Fe supercells with 54 lattice sites and Brillouin zone sampling of $6\times6\times6$ $k$ points as well as for supercells with 128 sites and $3\times3\times3$ $k$ points. For test purposes supercells with 128 sites and $4\times4\times4$ $k$ points were considered. After introduction of a defect dimer, trimer, or quadromer into the supercell two types of calculations were performed: (i) The positions of atoms were relaxed at constant volume and shape of the supercell (constant volume calculations - CV). The supercell volume corresponds to that of a cell containing perfect bcc Fe. (ii) The positions of atoms as well as the volume and shape of the supercell were relaxed so that the total stress/pressure on the supercell became zero (zero pressure calculations - ZP). The accuracy of the relaxation calculations is determined by two criteria: (i) CV and ZP are stopped if the residual force acting on any atom falls below a given threshold, and (ii) at each step of CV and ZP the energy minimization is performed until the total energy change falls below another threshold. In the present work the threshold values were equal to or lower than $10^{-3}$ eV/Å and $10^{-5}$ eV, in first and the second case, respectively. In many cases, especially if the contributions of phonon and electron excitations to the free energy shall be determined, the thresholds $10^{-4}$ eV/Å and $10^{-7}$ eV were used. This ensures a high precision of the results.

The binding energy of an embedded defect cluster consisting of $n$ species (point defects or foreign atoms) is defined by

$$E_{bind} = E(X_1 + X_2 + ... + X_n) + (n-1)E_0 - \sum_{i=1}^{n} E(X_i) \qquad (1)$$

$E(X_1 + X_2 + ... + X_n)$ and $E(X_i)$ denote the total energy of supercells with the cluster $X_1 + X_2 + ... + X_n$ and the monomer of species $X_i$, respectively, while $E_0$ is the total energy of a supercell with perfect bcc Fe. By definition the value of $E_{bind}$ is negative if attraction between the species dominates. In the present work the data for the total energy of supercells with single defect species (v, C, N, O, Al, Si, Ti, V, Cr, Mn, Co, Ni, Cu, Y, Mo, W) obtained in Ref. [17] were used in order to calculate the quantity $E_{bind}$ for the defect clusters.

## 2.2. Contributions of phonon and electron excitations to the free energy

Vibrational frequencies of supercells with a defect dimer, trimer, or quadromer were calculated using the method implemented in the VASP code. This procedure employs the frozen phonon approach and the harmonic approximation (cf. Refs. [3,45]). In order to calculate the dynamical matrix, finite differences were used, i.e. each atom is displaced along each Cartesian coordinate, and from the forces the matrix is determined. Only symmetry inequivalent displacements are considered, and the remainder of the dynamical matrix is filled using symmetry considerations. In the present work each atom was displaced along each axis by a small positive and negative displacement with an absolute value of 0.015 Å. Within this framework the vibrational free energy $F^{vib}(T)$ of a supercell with $N$ atoms is given by

$$F^{vib}(T) = \sum_{i=1}^{3N-3}\left[\frac{1}{2}\hbar\omega_i + k_B T \ln\left(1-e^{-\frac{\hbar\omega_i}{k_B T}}\right)\right] \quad (2)$$

$$F^{vib}(T) = U^{vib}(T) - T S^{vib}(T) \quad (3)$$

where $\omega_i$ are the phonon frequencies, $U^{vib}(T)$ and $S^{vib}(T)$ denote the vibrational contribution to internal energy and entropy ($S^{vib}(T) = -\partial F^{vib}/\partial T$), respectively, and $k_B$ is the Boltzmann constant.

The vibrational contribution to the free binding energy of an embedded defect cluster is determined similarly to Eq. (1)

$$F_{bind}^{vib} = F^{vib}(X_1+X_2+...+X_n) + (n-1)F_0^{vib} - \sum_{i=1}^{n}F^{vib}(X_i) \quad (4)$$

$F^{vib}(X_1+X_2+...+X_n)$ and $F^{vib}(X_i)$ denote the vibrational free energy of supercells with the cluster $X_1+X_2+...+X_n$ and the monomer $X_i$, respectively, whereas $F_0^{vib}$ is the vibrational free energy of the perfect bcc-Fe supercell. In the following the data determined in Ref. [17] for the vibrational free energy of supercells with defect monomers (v, C, N, O, Al, Si, Ti, V, Cr, Mn, Co, Ni, Cu, Y, Mo, W) and with perfect bcc Fe are employed to calculate the phonon contribution to the free binding energy of defect dimers, trimers, and quadromers.

In general two types of calculations may be performed: The dynamical matrix with the force derivatives and the vibrational frequencies are determined using ground-state atomic positions obtained either by CV or ZP. For point defects (monomers) an approximate relation between the CV- and ZP-based vibrational frequencies was recently found [17]

$$\omega_i^{ZP} \approx \omega_i^{CV}\left(1 - \gamma\frac{V-V_0}{V_0}\right) \quad (5)$$

where $V_0$ and $V$ are the supercell volumes in the CV and the ZP case, respectively, and $\gamma \approx 1.74$ is the thermodynamic or phonon Grüneisen parameter of bcc Fe. It was shown that relation (5) is valid if the relative arrangement of atoms in the environment of the defect is nearly identical in both types of calculations so that the difference between the frequencies is predominantly due to the volume change of the whole supercell in the ZP case. It is worth mentioning that Eq. (5) is formally very similar to the quasi-harmonic approach (cf. Eq. (7) and Refs. [46,47]). However, in the present case the volume difference is obtained from ground-state supercell volumes. Using relation (5) CV- and ZP-based vibrational free energies of the supercell with the defect can be converted into each other

$$F^{\text{vib,ZP}}(T) - F^{\text{vib,CV}}(T) = -\gamma \frac{V - V_0}{V_0} \left[ \sum_i \frac{1}{2} \hbar \omega_i^{CV} + \sum_i \hbar \omega_i^{CV} \left( e^{\frac{\hbar \omega_i^{CV}}{k_B T}} - 1 \right)^{-1} \right] \quad (6)$$

since the second term in Eq. (5) is usually relatively small, i.e. $(V - V_0)/V_0 \ll 1$ (cf. Table 2). In the present work it is investigated whether such a transformation is also applicable for the defect clusters.

Since $V$ is the volume of the supercell with a defect in the ground state under zero pressure conditions, the related ZP-based free energies are not exactly valid for $T \neq 0$ at zero pressure. However, the ZP-based phonon frequencies can be modified according to the quasi-harmonic approach (cf. e.g. Refs. [17,46,47])

$$\omega_i^{ZP,qh} = \omega_i^{ZP} \left( 1 - \gamma \frac{\Delta V}{V} \right) \quad (7)$$

Using this relation, frequencies can be determined for zero-pressure conditions at higher temperatures than the purely harmonic approach. The application of Eq. (7) is limited to temperatures where anharmonic effects do not prevail. The volume increase $\Delta V$ due to thermal expansion and due to the small effect of zero point vibrations is given by [46]

$$\frac{\Delta V}{V} = \frac{1}{B_0 V} \sum_{i=1}^{3N-3} \frac{1}{2} \hbar \omega_i^{ZP} + \beta T \tag{8}$$

In this equation $\beta$ and $B_0$ are the volume expansion coefficient and the bulk modulus of bcc Fe, respectively, with $\beta = 0.00004$ K$^{-1}$ and $B_0 = 170$ GPa [48,49]. Since the majority of experiments is performed at zero external pressure and elevated temperatures the defect free energy resulting from the quasi-harmonic correction to the ZP-based data is the most interesting quantity for a comparison. In the present paper the quasi-harmonic approach [Eq. (7)] is used under the assumption that the volume expansion $\Delta V$ of a supercell with a defect can be determined in a similar manner as for a supercell containing bulk bcc Fe and that the values of $\gamma$ are identical for both types of supercells.

The contribution of electronic excitations to the free energy of a supercell is calculated as in Ref. [17]

$$F^{el}(T) = U^{el}(T) - T S^{el}(T) \tag{9}$$

with the internal energy

$$U^{el}(T) = \sum_{i=1}^{2} \left[ \int_{\varepsilon_F}^{\infty} n_i(\varepsilon)(\varepsilon - \varepsilon_F) f(\varepsilon,T,\mu) d\varepsilon + \int_{0}^{\varepsilon_F} n_i(\varepsilon)(\varepsilon_F - \varepsilon)[1 - f(\varepsilon,T,\mu)] d\varepsilon \right] \tag{10}$$

and the entropy

$$S^{el}(T) = -k_B \sum_{i=1}^{2} \int \left[ f(\varepsilon,T,\mu) \ln f(\varepsilon,T,\mu) + (1 - f(\varepsilon,T,\mu)) \ln(1 - f(\varepsilon,T,\mu)) \right] n_i(\varepsilon) d\varepsilon \tag{11}$$

where $n(\varepsilon)$ and $f(\varepsilon,T,\mu)$ are the electronic density of states (DOS) in the ground state and the Fermi-Dirac distribution function, respectively, while $\varepsilon_F$ stands for the Fermi energy. In Eqs. (10-11) the sum is over the two spin orientations. The quantity $\mu(T)$ denotes the chemical potential of the electrons which is implicitly given by

$$N^{el} = \sum_{i=1}^{2} \int n_i(\varepsilon) f(\varepsilon,T,\mu) d\varepsilon \tag{12}$$

with $N^{el}$ as the total number of electrons.

The electronic contribution to the free binding energy of the defect clusters is defined by a relation similar to Eq. (4). In present calculations the corresponding data from Ref. [17] for the electronic free energy of supercells containing single defects and perfect bcc Fe are used.

## 3. Results and discussion

### *3.1. Defect clusters in the ground state*

Table 1 shows the calculated binding energy $E_{bind}$ of defect pairs in bcc Fe consisting of a vacancy and a solute ($2p$:C, N, O; $3p$: Al, Si; $3d$:Ti, V, Cr, Mn, Co, Ni, Cu; $4d$:Y, Mo; $5d$:W), as well as $E_{bind}$ of vacancy and Cu dimers, trimers and quadromers. The most favorable position of C, N, and O atoms in bcc Fe is the octahedral interstitial site while the other solutes prefer regular lattice sites, cf. Ref. [17]. Therefore, in the case of v-C, v-N, and v-O pairs the calculation was started with an initial configuration, where the distance between the two species is half of the lattice constant of bcc Fe, i.e. the vacancy on a bcc lattice site, and C, N, or O on an octahedral interstitial site. The initial distance between the constituents of the other dimers was equal to the first neighbor distance. Additionally, v-v, Cu-Cu, and v-Cu pairs with an initial distance equal to the second neighbor distance were considered [v-Cu(2), v$_2$(2), Cu$_2$(2)]. The initial configurations of the defect trimers (v$_3$, Cu$_3$) and quadromers (v$_4$, Cu$_4$) studied in this work are shown in Fig. 1. It was found that in the ground state these trimers and quadromers are the most stable amongst all possible geometrical arrangements. In most cases the relaxation of the supercell with the defect cluster under constant volume and zero pressure conditions leads to slight changes of atomic arrangements within and in the vicinity of the cluster. An exception is the v-Y pair where the relaxation shifts the Y atom to a position in the middle between the bcc sites initially occupied by v and Y [27,61]. The values of the binding energy of defect dimers listed in Table 1 vary considerably. The strongest and weakest bond is found for the v-O pair and for the Cu dimer at the second neighbor distance, respectively. The only repulsive interaction is found for the v-Co pair. However, in this case the value of $E_{bind}$ is very small so that the two parts of the dimer do not interact significantly. In Table 1 DFT data from the literature are shown as well. Most of them were obtained by CV. It should be mentioned that the majority of the literature data was obtained using VASP, except those of Refs. [50,52,56,66] (SIESTA code), [51] (CASTEP code), as well as those of Ref. [56] marked by the superscript "c" (PWSCF code). Only those VASP data from literature are shown that were determined using pseudopotentials generated by the PAW approach. For transition metals and magnetic systems the PAW method is generally preferred compared to the formerly used ultra-soft pseudopotential (USPP) approach [53,63,67]. In general the agreement between the literature data and the results of the present work is good. Larger differences are found by comparing with the SIESTA data [50,52,66] for the v-C, v-N, and v-O pairs, and for the $v_4$ quadromer. It must be noted that Terentyev *et al.* [68] emphasized that the absolute value of the binding energy of the v-C dimer should be at least 0.64 eV, which is in contrast to the SIESTA result. The reason for the discrepancy with VASP results of Refs. [63] (v-Mn), [59] and [58] (v-O), as well as [53] (v-Mo) is not clear. A possible cause might be the lower plane wave cutoff and/or the smaller $k$-point grid used in these studies. Table 1 shows that the data obtained for a supercell with 128 bcc sites are rather similar to those for a supercell with 54 sites, both for CV and ZP. This is an indication that in the case of defect dimers the choice of a supercell with 54 bcc sites already leads to reasonable data on ground state energetics. Test calculations showed that for defect trimers and quadromers supercells with at least 128 bcc sites should be chosen. Moreover, for a given supercell size results of CV and ZP are often very similar. In contrast to the results for the formation energy of a single vacancy and single solutes [17] there is no general trend for the binding energies of the defect clusters given in

Table 1: In some cases the CV data are slightly higher than the ZP data, in other cases it is the other way around. For a discussion on the relation between CV and ZP results in dependence on the size of the supercell and on their convergence, the reader is referred to former studies, cf. Refs. [17,69].

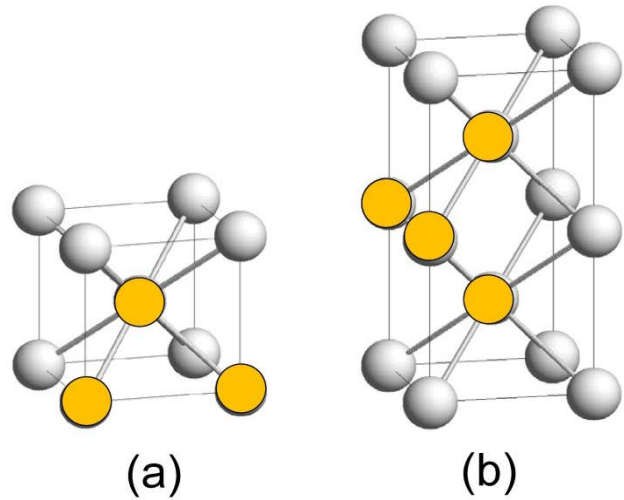

**Fig. 1.** Configuration of the defect trimer (a) and quadromer (b) before the relaxation under constant volume and zero pressure conditions. The positions of the defect (vacancy or Cu) are marked by yellow spheres.

In the case of CV the existence of a defect cluster in the supercell leads to the buildup of an internal pressure. If the ZP procedure is applied the volume of the supercell expands or contracts depending on the type of the defect cluster. The corresponding data for pressure and lattice parameter are given in Table 2. If the ZP method is used, amongst the vacancy-solute pairs v-Si leads to the strongest contraction of the supercell, whereas v-Y leads to the largest expansion. Both results are consistent with the finding that for a single vacancy and a single Si atom a contraction is found, and that a single Y atom leads to an expansion larger than the contraction due to a single vacancy [17,70]. For vacancy dimers, trimers, and quadromers a contraction is found while Cu dimers, trimers and quadromers cause an expansion.

Interestingly, in these cases the contraction/expansion per cluster constituent decreases with increasing size of the cluster. This may be related to the trend found for the binding energy per cluster constituent derived from the data shown in Table 1: The absolute value of this quantity increases with cluster size. As expected, the internal pressure $p$ and the relative change of the cell volume $(V-V_0)/V_0$ decrease if the supercell size increases. On the other hand, in many cases the absolute volume difference $(V-V_0)$ does not change much with supercell size, which is another indication that a cell with 54 lattice sites is often an acceptable choice. For certain defect clusters the volume change is combined with an isotropic expansion or contraction of the supercell. In other cases tetragonal and orthorhombic distortions are found. The change of the supercell shape is determined by the symmetry of the respective defect cluster.

### *3.2. Phonon and electron contributions to the free binding energy of the clusters*
### *3.2.1. Vacancy-solute dimers*

Figs. 2-4 depict the vibrational contribution $F_{\text{bind}}^{\text{vib}}$ and the sum of the vibrational and electronic contributions $F_{\text{bind}}^{\text{vib}}+F_{\text{bind}}^{el}$ to the free binding energy of vacancy-solute dimers by thick and thin lines, respectively. CV- and ZP-based data are shown. Note that the nonzero value of $F_{bind}^{\text{vib}}$ at $T=0$ is due to the zero point vibrations. It was verified that in most cases the CV- and the ZP-based data for $F_{\text{bind}}^{\text{vib}}$ can be transformed into each other by applying Eq. (6) to the supercell with the defect dimer and to the two supercells with the corresponding point defect. For the v-C, v-N, v-Mo, and v-Y dimers this transformation does not work so well. Here the difference between CV- and ZP-based data concerning the relative arrangement of atoms in the defect region is larger than in the case of the other vacancy-solute dimers. From detailed investigations one may conclude that a transformation between CV- and the ZP-based data for $F_{\text{bind}}^{\text{vib}}$ yields correct results as long as the difference between the corresponding atomic positions is less than about 0.02–0.04 Å, for the majority of atoms in the supercell. Furthermore, it was found that the use of the quasi-harmonic correction (7) to the ZP-based data does not lead to significant changes of the corresponding curves. Therefore, these data are generally valid in a temperature range where anharmonic effects are not dominant. Another limitation is given by the magnetic properties. In present calculations is was assumed that the magnetization of bulk iron corresponds to its value at $T = 0$. On the other hand the ferromagnetic-to-paramagnetic transition occurs at 1043 K. Furthermore, at 1183 K the $\alpha$-to-$\gamma$-phase transition takes place in iron. Therefore, in the high temperature range the curves presented must be considered with care, since in the calculation of these data the above mentioned effects were not taken into account.

In the examples of Figs. 2-4 the ZP-based data for $F_{\text{bind}}^{\text{vib}} + F_{\text{bind}}^{el}$ are generally higher than the data obtained by CV, i.e. under zero pressure conditions these defect dimers are less stable than under constant volume conditions. $F_{\text{bind}}^{\text{vib}}$ and $F_{\text{bind}}^{\text{vib}} + F_{\text{bind}}^{el}$ can increase or decrease with temperature, which means a decrease or increase of the stability of the vacancy-solute pair. In order to quantify this behavior the ratio of the sum of phonon and electron contributions $F_{\text{bind}}^{\text{vib}} + F_{\text{bind}}^{el}$ at 1000 K and the binding energy $E_{bind}$ in the ground state is given in Table 3. A positive/negative value corresponds to an increase/decrease of the absolute value of the total free binding energy $E_{bind} + F_{\text{bind}}^{\text{vib}} + F_{\text{bind}}^{el}$. In the case of the ZP-based data, which are most relevant for practical applications, the largest decrease of this value is found for the v-W pair (43%) while the largest increase is obtained for the v-Mn (35%). Also in many other cases the absolute value of the ratio $(F_{\text{bind}}^{\text{vib}} + F_{\text{bind}}^{el})/E_{bind}$ is above 20%. This clearly demonstrates that contributions of phonon and electron excitation to the free binding energy of the defect dimers must be taken into account, i.e. in most examples the ground state value $E_{bind}$ does not adequately describe the defect dimers at elevated temperatures. Due to the interplay of the different quantities in Eq. (4) and the corresponding relation for $F_{bind}^{el}$ the contribution of phonon and electron excitations to the free binding energy can be negative or positive, and lead to mostly monotonously increasing or decreasing functions of temperature. The latter also holds for the sum $F_{bind}^{vib} + F_{bind}^{el}$, with the exception of the ZP-based curves for v-Cr, v-Ni, and v-Y. The values of $F_{bind}^{el}$ can amplify the deviation of the free binding energy from the ground state value, which are caused by phonon excitations, or can reduce it (cf. Table 3). The sum $F_{bind}^{vib} + F_{bind}^{el}$ is a nonlinear function of $T$. At temperatures above about 500 - 600 K the derivative of $F_{bind}^{vib}$, i.e. the corresponding entropy, becomes a constant. This can be explained by considering Eq. (2) in the high-temperature limit $\hbar\omega_i << k_B T$, in combination with Eq. (4). In contrast, $F_{bind}^{el}$ and, therefore, also $F_{bind}^{vib} + F_{bind}^{el}$ has not a constant derivative at high temperatures.

    Because of the high computational effort to determine the vibrational frequencies for the supercell with the defect dimer and for the two supercells containing the corresponding point defect, only a few calculations were performed to obtain phonon and electron contributions to the free binding energy for a system with 128 bcc sites. At this point it should be also mentioned that the symmetry of the supercell with a defect dimer is often lower than in the case of a point defect. This leads to an increased computing time for the determination of the vibrational frequencies compared to the same task for the supercell with a point defect. In the case of the v-C, v-N, v-O, v-Cu, and v-Cu(2) pairs the a satisfactory agreement of the ZP-based data for a system with 128 bcc sites with those obtained for a supercell with 54 bcc sites is found. At 1000 K the difference between the results for $F_{bind}^{vib} + F_{bind}^{el}$ is less than about 0.04 eV. For the v-Cu pair additional calculations with

$4 \times 4 \times 4$ $k$ points and 128 bcc sites were performed and at 1000 K a shift of +0.01 eV was found compared to results obtained using $3 \times 3 \times 3$ $k$ points.

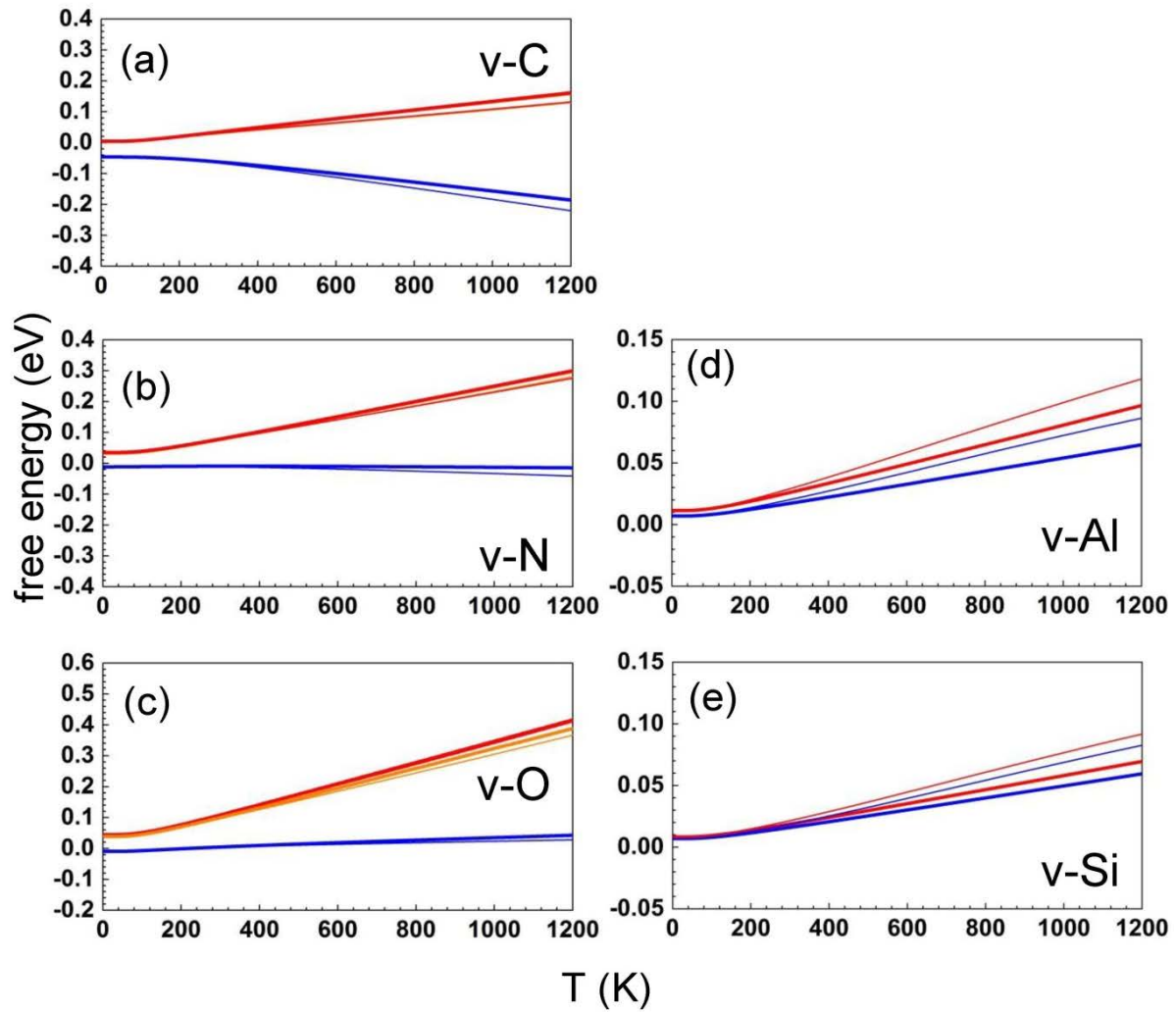

**Fig. 2**. Vibrational contribution $F_{bind}^{vib}$ (thick lines) and the sum of the vibrational and electronic contributions $F_{bind}^{vib} + F_{bind}^{el}$ (thin lines) to the free binding energy of vacancy-$2p$-solute and vacancy-$3p$-solute pairs in bcc Fe. Note that in a few cases thick and thin lines lie close together. CV-based and ZP-based data obtained for a supercell with 54 bcc sites are depicted by blue and red lines, respectively. ZP-based data for a supercell with 128 bcc sites are shown in orange color. For v-C and v-N the red and orange lines are nearly identical.

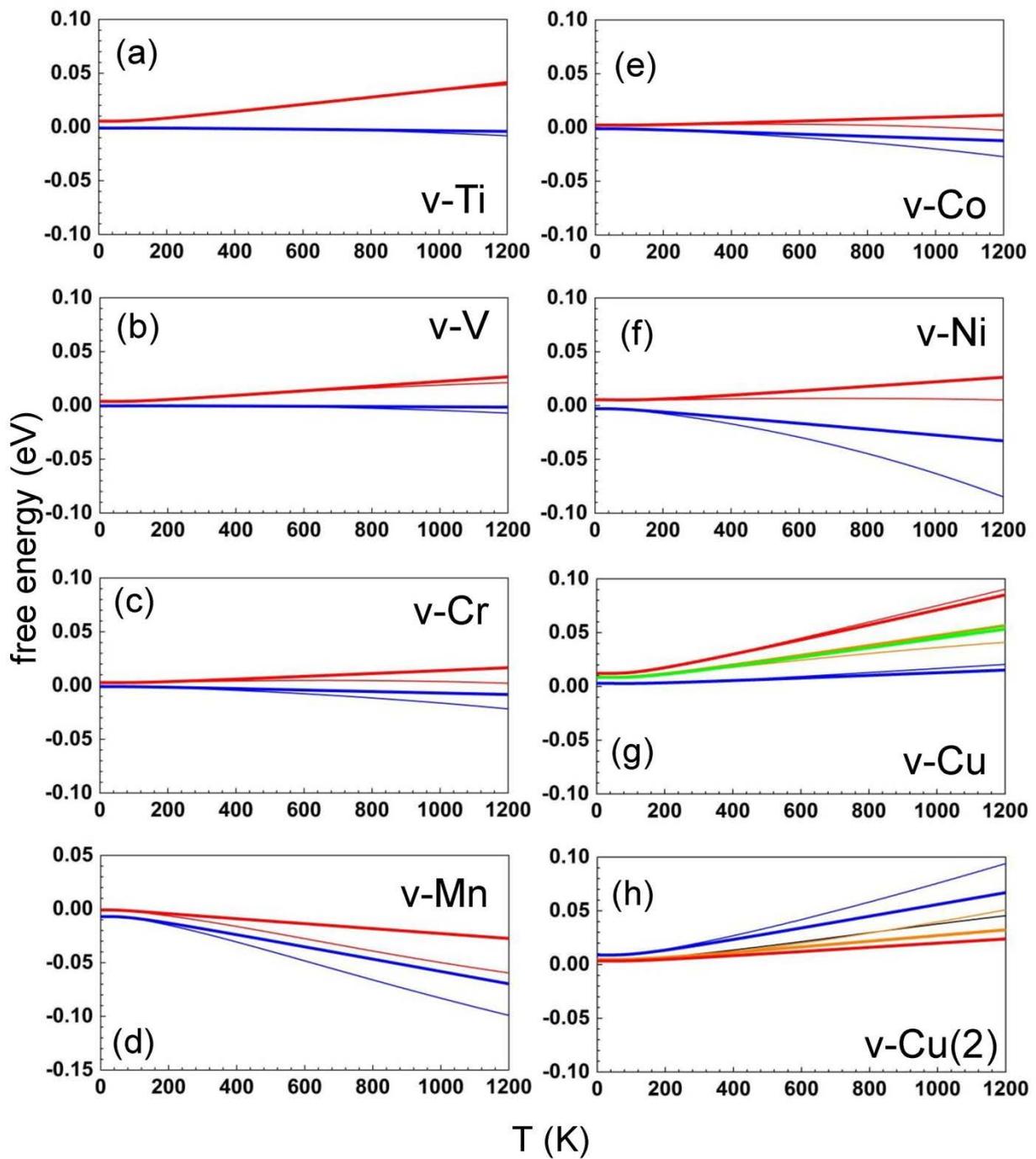

**Fig. 3**. Vibrational contribution $F_{bind}^{vib}$ and the sum of the vibrational and electronic contributions $F_{bind}^{vib} + F_{bind}^{el}$ to the free binding energy of vacancy-$3d$-solute dimers in bcc Fe. The green line in (g) shows ZP-based results of additional calculations with $4 \times 4 \times 4$ $k$ points and 128 bcc sites. The meaning of the other lines and colors is explained in the caption of Fig. 2.

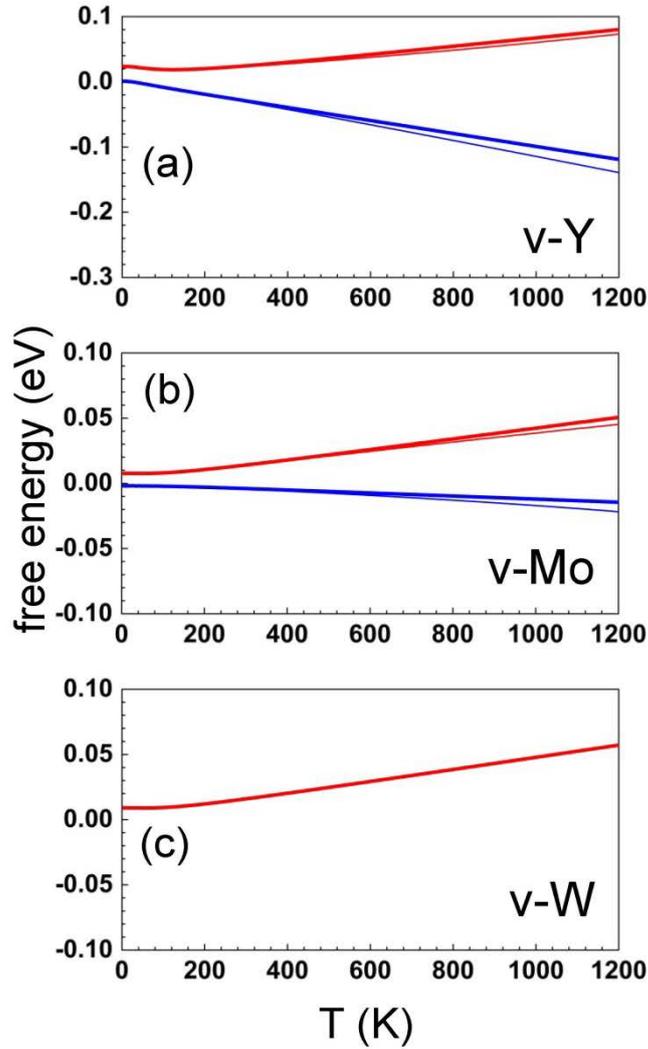

**Fig. 4**. CV- and ZP-based data of $F_{bind}^{vib}$ and $F_{bind}^{vib} + F_{bind}^{el}$ for vacancy- $4d$ -solute and vacancy- $5d$ -solute pairs in bcc Fe. The style of the presentation is identical to that in Figs. 2 and 3.

*3.2.2. Vacancy and copper dimers, trimers and quadromers*

For small vacancy and Cu clusters in bcc Fe the temperature dependence of the ZP-based data for $F_{bind}^{vib}$ and $F_{bind}^{vib} + F_{bind}^{el}$ is shown in Figs. 5 and 6. While for dimers supercells with both 54 and 128 bcc sites were considered in the calculations, only 128 bcc sites were used in the case of trimers and quadromers. Compared to the case of defect dimers the computational effort to calculate the vibrational frequencies for trimer and quadromers is often higher due to the lower symmetry of the defect arrangement in the supercell. The general behavior of $F_{bind}^{vib}$ and $F_{bind}^{vib} + F_{bind}^{el}$ for these defect clusters is similar to that of the vacancy-solute pairs. The corresponding characteristics are summarized in Table 3. In the case of the dimers in the smaller supercell, $v_2$ and $Cu_2$ lose their stability with increasing temperature whereas the opposite is found for $v_2(2)$ and $Cu_2(2)$. For the supercell with 128 bcc sites and calculations with $3\times3\times3$ $k$ points the value of $F_{bind}^{vib} + F_{bind}^{el}$ increases/decreases for Cu/vacancy pairs. Obviously, a satisfactory convergence with respect to the size of the supercell could not be reached in these cases, although the difference between the results for the two supercell sizes is less than 0.06 eV (at 1000 K). For $v_2$ and $Cu_2$ calculations with $4\times4\times4$ $k$ points and 128 bcc sites were performed as well and shifts of +0.04 and +0.01 eV (1000 K) were found compared to results obtained using $3\times3\times3$ $k$ points. These results show the difficulties in such type of calculations where differences of hundredths of eV determine the qualitative behavior as the increase or decrease with temperature. Further investigations should be performed in order to clarify whether the differences found are due to real supercell size effects or caused by the limited precision of the methodology used in present DFT calculations. The stability of the vacancy clusters $v_3$ and $v_4$ decreases with temperature, at 1000 K by about 14 and 18%, respectively. In the case of $Cu_3$ and $Cu_4$ the decrease is even 36 and 45%, respectively. This is a clear indication that at elevated temperatures the contributions of phonon and electron excitation to the free binding energy of the defect trimers and quadromers is not negligible compared to the corresponding ground state value $E_{bind}$.

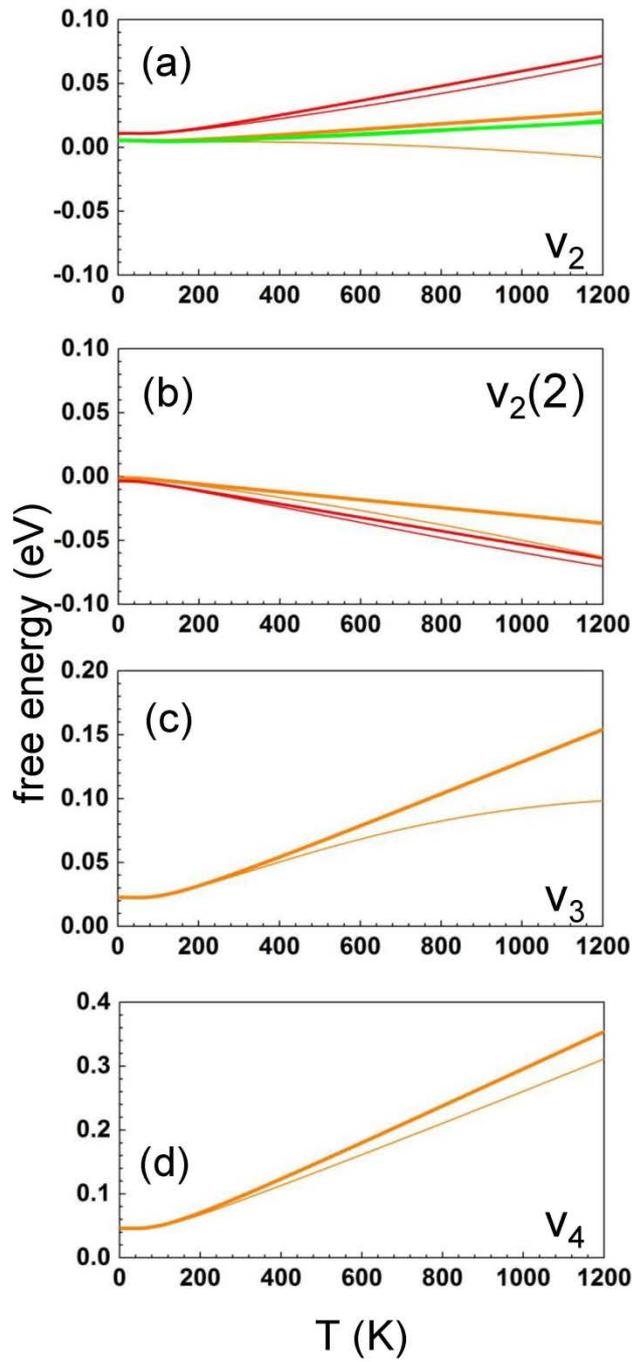

**Fig. 5.** Phonon contribution $F_{bind}^{vib}$ and the sum of phonon and electron contributions $F_{bind}^{vib} + F_{bind}^{el}$ to the free binding energy of small vacancy clusters in bcc Fe.

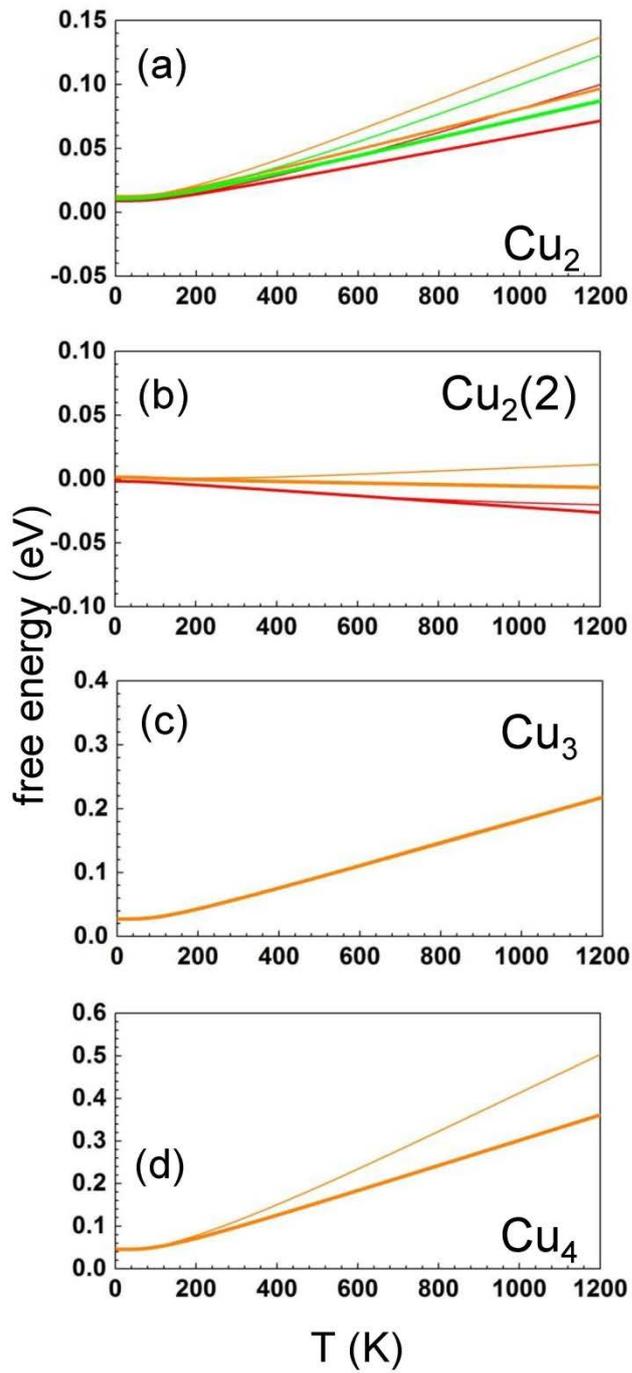

**Fig. 6**. CV- and ZP-based data of $F_{\text{bind}}^{\text{vib}}$ and $F_{\text{bind}}^{\text{vib}} + F_{\text{bind}}^{el}$ for small Cu clusters in bcc Fe.

## 4. Summary and conclusions

The ground-state binding energy of vacancy-solute pairs and small vacancy and Cu clusters in bcc Fe was determined considering atomic relaxations under constant volume (CV) and zero pressure (ZP) conditions and supercells with 54 and 128 lattice sites. The comparison between results determined by CV and ZP and for the two different supercell sizes indicates that in many cases the choice of a supercell with 54 bcc sites leads to reasonable values. Furthermore, a good agreement with DFT data from literature is found.

Based on the equilibrium atomic positions determined in the ground state by both CV and ZP, the contribution of phonon excitations to the free binding energy was calculated within the framework of the harmonic approximation. The contribution of electron excitation was obtained using the corresponding ground state data for the electronic density of states. At elevated temperature the vacancy-solute pairs considered in this work are less stable under zero pressure than under constant volume conditions. It was found that the quasi-harmonic correction to the ZP-based data does not lead to significant changes of the corresponding curves. Therefore, these data are valid under zero-pressure conditions at higher temperatures than in the framework of the purely harmonic approach. Such conditions are usually realized in experiments and practical applications. However, the validity of present results is limited if anharmonic effects and magnon excitations become dominant. The phonon contribution $F_{\text{bind}}^{\text{vib}}$ and the sum of phonon and electron contributions $F_{\text{bind}}^{\text{vib}} + F_{\text{bind}}^{el}$ to the free binding energy can increase or decrease with temperature, which means a decrease or increase of the stability of the vacancy-solute dimer. A quantification of this relatively complex behavior was performed using the ratio of $F_{\text{bind}}^{\text{vib}} + F_{\text{bind}}^{el}$ to the ground state binding energy $E_{bind}$ at 1000 K. In the case of the ZP-based data, which are most relevant for practical applications, the largest decrease of the absolute value of the free binding energy is found for v-W (43%), and the largest increase is obtained for v-Mn (35%). Present results are a clear indication that contributions of phonon and electron excitation to the free binding energy of the defect dimers cannot be neglected, i.e. in most examples the ground state value $E_{bind}$ does not describe correctly the dimers at elevated temperatures. The contribution of electron excitations $F_{bind}^{el}$ can increase the deviation of the free binding energy from the ground state value, which is caused by phonon excitations, or can compensate it. In all cases the sum $F_{bind}^{vib} + F_{bind}^{el}$ is a nonlinear function of $T$ which is due to the temperature dependence of $F_{bind}^{el}$. For v-solute pairs the ZP-based results for a system with 128 bcc sites agree satisfactorily with those obtained for a supercell with 54 bcc sites. The general behavior of $F_{bind}^{vib}$ and $F_{bind}^{vib} + F_{bind}^{el}$ for small vacancy and Cu clusters is similar to that of the vacancy-solute pairs. However, the dependence of the results on the supercell size is more sensitive. The reason for this should be investigated in more detail. For most of the defect clusters it was found that the CV- and the ZP-based data for $F_{bind}^{vib}$ can be transformed into each other using a recently derived relation [17]. As shown in detail earlier [17], the CV- and ZP-based values of vibrational contribution to the total free energy of a supercell with a defect do not become equal with increasing supercell size. This is in contrast to the size convergence obtained in the case of ground state energetics [17,69].

The results obtained in this work are of general importance for studies on the thermodynamics and kinetics of defect clusters in solids: First, the methodology used to determine the free energy of clusters in bcc Fe can be applied to other solids in order to estimate the influence of phonon and electron excitations on the free energy of defect clusters. Second, the results on the temperature dependence of the free binding energy of vacancy-solute pairs may contribute to an improved modeling of the diffusion of foreign atoms in bcc Fe by the vacancy mechanism. For this purpose, more ZP-based data for pairs with distances between the constituents greater than the first neighbor distance should be determined in forthcoming studies. A similar methodology may be utilized in order to improve the modeling of solute diffusion by the dumbbell mechanism. Third, the free binding energy of defect dimers determined in this work may be used to get temperature-dependent parameters for the description of pair interactions within the framework of rigid-lattice kinetic and Metropolis Monte Carlo simulations. These data can replace the presently employed values obtained from ground state calculations. In this manner the accuracy of modeling may be improved considerably. Also in this case an extension of the data base by considering more temperature-dependent second-neighbor pair interaction parameters should be envisaged.

**Acknowledgements**

This work contributes to the Joint Programme on Nuclear Materials (JPNM) of the European Energy Research Alliance (EERA). It was partially funded by the European Horizon 2020 Research and Innovation Action Programme under Grant Agreement No. 661913 (SOTERIA).

The authors thank the Center for Information Services and High Performance Computing (ZIH) at TU Dresden for generous allocations of computer time. High Performance Computing facilities at Helmholtz-Zentrum Dresden - Rossendorf (HZDR) are also gratefully acknowledged.

**Table 1.** Binding energy data for vacancy-solute pairs, as well as vacancy and Cu dimers, trimers and quadromers, determined under constant volume (CV) and zero pressure (ZP) conditions, for two different supercell sizes, in comparison with DFT data from literature (superscripts: refererences). Before relaxation the distance between the vacancy and the octahedral interstitials C, N, or O was half of the lattice constant of bcc Fe, whereas the initial distance between the vacancy and the other solutes was equal to the first neighbor distance. Additionally, v-Cu, v-v, and Cu-Cu pairs with constituents in the second neighbor distance [v-Cu(2), $v_2$(2), $Cu_2$(2)] were considered. The initial configurations of the defect trimers ($v_3$, $Cu_3$) and quadromers ($v_4$, $Cu_4$) investigated in this work are shown in Fig. 1.

| Defect | $E_{bind}$ (eV) | | | | | |
|---|---|---|---|---|---|---|
| | CV | | Literature data | ZP | | Literature data |
| | 54 | 128 | | 54 | 128 | |
| v-C | -0.7113 | | -0.47[a] [50], -0.62[b] [51], -0.70 [54], -0.68 [55], -0.64 [57] | -0.5409 | -0.5839 | -0.41[a] [52], -0.59 [53], -0.58[b] [51], -0.52[a] [56], -0.65[c] [56] |
| v-N | -0.9739 | | -0.78[a] [50] | -0.8067 | -0.8619 | -0.75[a] [52], -0.86 [53] |
| v-O | -1.735 | -1.6710 | -1.45 [58], -1.55 [59], -1.65 [60], -1.69 [61], | -1.527 | -1.576 | -1.44[a] [52], -1.53 [17], -1.53 [22] |

|  |  |  | -1.74 [17], -1.52[a] [50] |  |  |  |
|---|---|---|---|---|---|---|
| **v-Al** | -0.3230 |  | -0.32 [24] | -0.3162 |  | -0.29 [26] |
| **v-Si** | -0.2848 |  | -0.30 [24], -0.32 [62] | -0.2880 |  | -0.29 [53] |
| **v-Ti** | -0.2528 |  | -0.25 [24], -0.24 [63], -0.26 [60], -0.23 [61], -0.29 [54] | -0.2367 |  | -0.22 [53] |
| **v-V** | -0.06274 |  | -0.056 [24], -0.047 [63], -0.06 [62] | -0.05097 |  | -0.04 [53] |
| **v-Cr** | -0.05733 |  | -0.057 [24], -0.056 [28], -0.061 [63], -0.05 [60], -0.045 [54] | -0.04500 |  | -0.05 [53] |
| **v-Mn** | -0.1525 |  | -0.17 [24], -0.17 [28], -0.20 [63], -0.17 [62], -0.16 [54] | -0.1410 |  | -0.16 [53] |
| **v-Co** | 0.01941 |  | 0.018 [24], 0.012 [63], | 0.02410 |  |  |
| **v-Ni** | -0.1040 | | -0.10 [24], -0.10 [28], -0.11 [64], -0.11 [54] | -0.08733 | | |
| **v-Cu** | -0.2541 | -0.2449 | -0.26 [24], -0.26 [28], -0.27 [63], -0.26 [54] | -0.2466 | -0.2421 -0.2393* | -0.24 [53] |
| **v-Cu(2)** | -0.1877 | -0.1761 | -0.17 [24], -0.17 [28], -0.16 [63], -0.17 [54] | -0.1737 | -0.1692 | |
| **v-Y** | -1.390 | | -1.45 [60], -1.26 [61] | -1.341 | | |
| **v-Mo** | -0.1669 | | -0.16 [24], -0.17 [63], -0.17 [3] | -0.1342 | | -0.33 [53] |
| **v-W** | | | -0.13 [24], -0.14 [63], -0.14 [62], -0.14 [3] | -0.1115 | | |
| **v₂** | --0.1540 | -0.1525 | -0.15 [64], -0.16 [65] | -0.1696 | -0.1550 -0.1618* | |
| **v₂(2)** | -0.1931 | -0.2252 | -0.21 [64], -0.23 [65] | -0.2143 | -0.2318 | |
| **v₃** | | -0.6592 | | | -0.6697739 | |
| **v₄** | | -1.405 | | | -1.415 | -1.21[a] [66] |
| **Cu₂** | -0.2249 | -0.2375 | -0.25 [64], -0.25 [54] | -0.2293 | -0.2415 -0.2329* | -0.18 [2] |
| **Cu₂(2)** | -0.04486 | -0.05138 | -0.060 [64], -0.058 [54] | -0.05368 | -0.05684 | |
| **Cu₃** | | -0.4939 | | | -0.5070 | |
| **Cu₄** | | -0.8968 | -0.80 [18] | | -0.9166 | |

*result of calculations with $4\times 4\times 4$ $k$ points

[a]SIESTA code, norm-conserving pseudopotentials

[b]CASTEP code, PAW

[c]PWSCF code, USPP pseudopotentials

**Table 2.** Internal pressure $p$ in the supercell with a defect cluster, obtained after relaxation under constant volume conditions (CV) as well as the lattice parameter $a$ and the volume change $V - V_0$ determined by relaxation under zero pressure (ZP) conditions. If CV yields a negative/positive pressure a volume decrease/increase is obtained in the case of ZP. Results for two different supercell sizes are shown. If the ZP procedure leads to an isotropic contraction/expansion only one line of data is given for the corresponding defect. In the case of tetragonal distortions the first line is related to the two equivalent directions (a) whereas the second line is related the third direction (b). In the case of an orthorhombic distortion three lines (a), (b), and (c) are given. The volume $V_0$ of the supercell with perfect bcc Fe can be determined using the lattice parameter $a_0 = 2.834$ Å (cf. Ref. [17]).

| Defect | CV $p$ (eV/Å³) | | ZP $a$ (Å) | | $V - V_0$ (Å³) | |
|---|---|---|---|---|---|---|
| | 54 | 128 | 54 | 128 | 54 | 128 |
| v-C | $-6.4028 \times 10^{-4}$ (a) | | 2.8254 (a) | 2.8295 (a) | 0.76631 | 0.41464 |
| | $5.0799 \times 10^{-3}$ (b) | | 2.8481 (b) | 2.8382 (b) | | |
| v-N | $6.0695 \times 10^{-4}$ (a) | | 2.8261 (a) | 2.8297 (a) | 1.5534 | 1.2584 |
| | $6.8598 \times 10^{-3}$ (b) | | 2.8504 (b) | 2.8394 (b) | | |
| v-O | $4.3312 \times 10^{-3}$ (a) | $1.6565 \times 10^{-3}$ (a) | 2.8350 (a) | 2.8344 (a) | 2.8757 | 2.6343 |
|  | $6.6482\times10^{-3}$ (b) | $2.7523\times10^{-3}$ (b) | 2.8441 (b) | 2.8381 (b) |  |  |
| **v-Al** | $-1.8253\times10^{-3}$ |  | 2.8320 |  | -1.0476 |  |
| **v-Si** | $-5.5600\times10^{-3}$ |  | 2.8285 |  | -3.2805 |  |
| **v-Ti** | $1.1333\times10^{-3}$ |  | 2.8347 |  | 0.70740 |  |
| **v-V** | $-7.4957\times10^{-4}$ |  | 2.8328 |  | -0.53190 |  |
| **v-Cr** | $-6.4350\times10^{-4}$ |  | 2.8329 |  | -0.47520 |  |
| **v-Mn** | $-1.5863\times10^{-3}$ |  | 2.8322 |  | -0.93690 |  |
| **v-Co** | $-3.4589\times10^{-3}$ |  | 2.8307 |  | -1.9035 |  |
| **v-Ni** | $-8.5299\times10^{-4}$ |  | 2.8324 |  | -0.7452 |  |
| **v-Cu** | $-2.1201\times10^{-3}$ | $-1.3679\times10^{-3}$ | 2.8310 | 2.8327<br>2.8324* | -1.6578 | -1.9008<br>-1.7728* |
| **v-Cu(2)** | $2.0730\times10^{-3}$(a)<br>$2.0185\times10^{-3}$(b) | $2.9504\times10^{-4}$(a)<br>$4.7768\times10^{-4}$(b) | 2.8349 (a)<br>2.8347 (b) | 2.8339 (a)<br>2.8346 (b) | 0.8035 | 0.30431 |
| **v-Y** | $1.5654\times10^{-2}$ |  | 2.8471 |  | 8.9316 |  |
| **v-Mo** | $4.2426\times10^{-3}$ |  | 2.8372 |  | 2.3247 |  |
| **v-W** | | | 2.8357 | | 2.5569 | |
| **v₂** | $-7.9898 \times 10^{-3}$ | $-3.9613 \times 10^{-3}$ | 2.8251 | 2.8305<br>2.8302* | -5.5107 | -5.2928<br>-5.1392* |
| **v₂(2)** | $-8.0335 \times 10^{-3}$(a)<br>$-1.0793 \times 10^{-2}$(b) | $-3.9060 \times 10^{-3}$(a)<br>$-5.1966 \times 10^{-3}$(b) | 2.8279 (a)<br>2.8174 (b) | 2.8317 (a)<br>2.8274 (b) | -5.9476 | -5.7347 |
| **v₃** | $-1.0311 \times 10^{-2}$(a)<br>$-1.2539 \times 10^{-2}$(b) | $-5.1166 \times 10^{-3}$(a)<br>$-5.8867 \times 10^{-3}$(b) | | 2.8299 (a)<br>2.8274 (b) | | -7.6202 |
| **v₄** | | $-6.3370 \times 10^{-3}$(a)<br>$-5.9223 \times 10^{-3}$(b)<br>$-5.9048 \times 10^{-3}$(c) | | 2.8272 (a)<br>2.8288 (b)<br>2.8289 (c) | | -8.7819 |
| **Cu₂** | $8.7137 \times 10^{-3}$ | $3.1959 \times 10^{-3}$ | 2.8409 | 2.8367<br>2.8365* | 4.7520 | 4.2240<br>4.5376* |
| **Cu₂(2)** | $9.5884 \times 10^{-3}$(a)<br>$1.0805 \times 10^{-2}$(b) | $3.6423 \times 10^{-3}$(a)<br>$4.3526 \times 10^{-3}$(b) | 2.8403 (a)<br>2.8458 (b) | 2.8364 (a)<br>2.8388 (b) | 5.5963 | 4.9916 |
| **Cu₃** | $1.3138 \times 10^{-2}$(a) | $4.8305 \times 10^{-3}$(a) | | 2.8376 (a) | | 6.4860 |

|  | $1.4452 \times 10^{-2}$(b) | $5.3705 \times 10^{-3}$(b) |  | 2.8394 (b) |  |  |
|  |  |  |  |  |  |  |
| **Cu$_4$** |  | $6.3448 \times 10^{-3}$(a) |  | 2.8386 (a) |  | 7.5567 |
|  |  | $6.0413 \times 10^{-3}$(b) |  | 2.8395 (b) |  |  |

*result of calculations with $4 \times 4 \times 4$ $k$ points

**Table 3.** The ratio of the sum of phonon and electron contributions to the free binding energy to the value of the ground state binding energy [$\Delta = (F_{bind}^{vib} + F_{bind}^{el})/E_{bind}$] at 1000 K, obtained from the CV- and ZP-based data shown in Figs. 2-6. Note that a negative/positive value of $\Delta$ means a relative decrease/increase of the absolute value of the free binding energy, or of the stability of the defect cluster. Furthermore, in the case of calculations for a supercell with 54 bcc sites, the sign of the difference between ZP- and CV-based data $\varepsilon = (F_{bind}^{vib,ZP} + F_{bind}^{el,ZP}) - (F_{bind}^{vib,CV} + F_{bind}^{el,CV})$ is given. The electronic contribution to the free binding energy $F_{bind}^{el}$ can amplify (A) the deviation of the total free binding energy from the ground state value, which are caused by phonon excitations ($F_{bind}^{vib}$), can reduce (R) this deviation, or is negligible (N).

| Defect | 54 | | | | | 128 | |
|---|---|---|---|---|---|---|---|
| | CV | | ZP | | | ZP | |
| | $\Delta$ in % | $F_{bind}^{el,CV}$ | $\Delta$ in % | $F_{bind}^{el,ZP}$ | $\varepsilon$ | $\Delta$ in % | $F_{bind}^{el,ZP}$ |
| v-C | +26 | A | -20 | R | > 0 | -18 | R |
| v-N | +3 | A | -28 | R | > 0 | -27 | R |
| v-O | -1 | R | -22 | N | > 0 | -21 | R |
| v-Al | -22 | A | -31 | A | > 0 | | |
| v-Si | -24 | A | -26 | A | > 0 | | |
| v-Ti | +2 | A | -14 | N | > 0 | | |
| v-V | +7 | A | -37 | R | > 0 | | |
| v-Cr | +29 | A | -4 | R | > 0 | | |
| v-Mn | +54 | A | +35 | A | > 0 | | |
| v-Co | +101 | A | +0.5 | R | > 0 | | |
| **v-Ni** | +60 | A | -6 | R | > 0 | | |
| **v-Cu** | -7 | A | -30 | A | > 0 | -15<br>-19* | A<br>N* |
| **v-Cu(2)** | -41 | A | -23 | A | < 0 | -22 | A |
| **v-Y** | +8 | A | -4 | R | > 0 | | |
| **v-Mo** | +10 | A | -29 | R | > 0 | | |
| **v-W** | | | -43 | N | | | |
| **v$_2$** | | | -31 | R | | +2<br>-10* | A<br>N* |
| **v$_2$(2)** | | | +28 | A | | +21 | A |
| **v$_3$** | | | | | | -14 | R |
| **v$_4$** | | | | | | -18 | R |
| **Cu$_2$** | | | -35 | A | | -47<br>-43* | A<br>A* |
| **Cu$_2$(2)** | | | +34 | R | | -16 | A |
| **Cu$_3$** | | | | | | -36 | N |
| **Cu$_4$** | | | | | | -45 | A |

*result of calculations with $4\times 4\times 4$ $k$ points

$V - V_0$.